\documentstyle[12pt,epsf]{ioplppt}
\jl{4}
\begin{document}
%
%
\title{\hfill\parbox{4cm}{\rm JINR E2-95-511\\
                    FAU-TP3-95/13\\
                    hep-ph/9512324}\\[10mm]Photoproduction of neutral pion pairs\\
       in the Coulomb field of the nucleus}[]

\author{
A A Bel'kov\dag\ftnote{3}{E-mail: {\tt belkov@cv.jinr.dubna.su}},
M Dillig\ddag~
and
A V Lanyov\dag
}

\address{\dag\ Particle Physics Laboratory,
 Joint Institute for Nuclear Research,
        141980 Dubna, Moscow Region, Russia}

\address{\ddag
Institut f\"ur Theoretische Physik III der Universit\"at Erlangen-N\"urnberg,
D-91058 Erlangen, Germany}

\begin{abstract}

   The total cross section for Coulomb photoproduction of neutral pion
pairs in the reaction $\gamma A\to\pi^0\pi^0 A$ is estimated within the
effective chiral lagrangian approach.
   The amplitude of $\gamma \gamma^{*} \to \pi^0 \pi^0$ with one
off-shell photon is calculated at $O(p^6)$ in the momentum expansion;
in addition, nuclear absorption is taken into account.
   Besides its experimental feasibility, the results of the calculation
demonstrate that the reaction $\gamma A\to\pi^0\pi^0 A$ is a powerful
source of information on the process $\gamma \gamma \to \pi^0 \pi^0$
close to threshold.

\end{abstract}
%
%
 \maketitle

\section{Introduction}
    The nucleus Coulomb interaction method is an effective approach
to study in high-energy experiments the low-energy electromagnetic
properties of pions in the process $\gamma\gamma\to\pi\pi$.
    There are two possible ways to realize the Coulomb interaction method
in experiment: the radiative scattering $\pi A \to\pi\gamma A$ of a pion
on a nucleus and the photoproduction of pion pairs in the Coulomb field
of a nucleus $\gamma A\to\pi\pi A$.
    The experiments under discussion are based on the fact that for
sufficiently small momentum transfers the interaction of high-energy
particle with nuclei is extremely peripheral and thus dominated by
scattering on the virtual photons of the Coulomb field of the nucleus.
    The reaction $\pi A \to\pi\gamma A$ allows to investigate the
$\gamma\gamma\to\pi\pi$ process in the region
$m^2_{\pi\pi}=(p_{\pi 1}+p_{\pi 2})^2<0$ and to measure the charged pion
polarizability \cite{Antipov} at the point $m^2_{\pi\pi}=0$, while the
reaction $\gamma A\to\pi\pi A$ occurs in the region
$m^2_{\pi\pi} \geq 4 m^2_\pi$ close to threshold.
    In the case of charged pions the two reactions give complementary
information about the process $\gamma\gamma\to\pi^{+}\pi^{-}$ in both
the physical and the non-physical region.
    In the case of neutral pions the photoproduction of a neutral pion
pair in the Coulomb field of a nucleus provides a new source of the
information on the process $\gamma\gamma\to\pi^0\pi^0$, which was
measured in the electron-positron experiment by the Crystal Ball
Collaboration \cite{crystal-ball} and is under active discussion in
the context of the physical programme at DA$\Phi$NE \cite{DAFNE1}.

    The present theoretical interest in the elementary process
$\gamma\gamma\to\pi^0\pi^0$ is caused both by experimental data
\cite{crystal-ball} mentioned above and recent progress in Chiral
Perturbation Theory (ChPT) \cite{Gasser1} up to and including $O(p^6)$.
    The process $\gamma\gamma\to\pi^0\pi^0$ is very sensitive to the
higher-order contributions in ChPT since the first non-vanishing
amplitude arises from meson loops at $O(p^4)$ without counterterms.
    However, the one-loop amplitude at $O(p^4)$ calculated in
paper \cite{Bijnens1} does not describe the data even near threshold
\cite{Bijnens3}.
    This is not surprising, as the analysis of the
$\gamma\gamma\to\pi^0\pi^0$ amplitude based on dispersion relations
\cite{dispersion} demonstrates the importance of
unitarity corrections corresponding to higher orders, next to $O(p^4)$.
    In fact, the two-loop calculation at $O(p^6)$ carried out in
paper \cite{Bellucci1} gives a considerable improvement of the
description within ChPT.
    For completeness we mention that there is also another
consideration of $\gamma\gamma\to\pi^0\pi^0$ up
to one-loop order corresponding to $O(p^5)$ in the treatment of
Generalized ChPT \cite{Knecht}.

    The contribution from the $O(p^6)$ counterterms was estimated in
paper \cite{Bellucci1}
\footnote{The same paper presents a diagrammatical expansion of
          the amplitude $\gamma\gamma\to\pi^0\pi^0$ up to  $O(p^6)$.}
from the low-energy meson phenomenology with
resonance exchange saturation.
    In papers \cite{Bellucci2,Belkov2} the counterterms were fixed
as effective meson lagrangians with higher-order derivative terms
obtained from the bosonization of Nambu--Jona-Lasinio model (NJL).
    The sensitivity of $\gamma\gamma\to\pi^0\pi^0$ to higher-order
corrections including Born contributions from the effective meson
lagrangian at $O(p^6)$ makes this process a valuable source of the
experimental information essential for the test of bosonized chiral
lagrangians at $O(p^6)$.

    In this paper we investigate the possibility of studying the process
$\gamma\gamma\to\pi^0\pi^0$ near threshold in the photoproduction of
neutral pion pairs in the Coulomb field of a nucleus.
    For the first time the reaction $\gamma A\to\pi\pi A$ was considered
in this context in paper \cite{Belkov1} with a one-loop amplitude of
$\gamma\gamma\to\pi^0\pi^0$ at $O(p^4)$.
    The total cross sections $\gamma A\to\pi\pi A$ were estimated
for energies of the incident photon at 20 GeV and 40~GeV for momentum
transfer cutoffs 5 MeV and 10~MeV, respectively.
    However, due to the large incident energy and small cutoff,
the nuclear absorption and the offshellness of the virtual photon
from the nuclear Coulomb field were not taken into account.

    In this note, we extend and improve the previous calculation
\cite{Belkov1} in a consistent way: we presents results of a
calculation of the nuclear Coulomb photoproduction of neutral pion
pairs, where we include in the elementary amplitude
$\gamma \gamma^{*}\to\pi^0\pi^0$ both off-shell corrections for the
Coulomb virtual photon $\gamma^{*}$ and contributions from ChPT up to
$O(p^6)$, and where, in addition, nuclear structure and absorption are
taken into account in the form factor for the
nucleus.

\section{Coulomb photoproduction}
     The photoproduction of a $\pi^0\pi^0$ pair in the Coulomb field of
a nucleus is schematically described by the diagram in figure 1.
     The virtual photon $\gamma^{*}(q_2)$ for the interaction of the
incident real photon with the stationary Coulomb field of the nucleus
has zero energy and transfers only momentum: $q_2 = (0,{\bf q}_2)$.
     Then the amplitude of the reaction
$\gamma(q_1) A \to \pi^0(p_1)\pi^0(p_2)A$ has the form
\begin{equation}
      T_{\rm C} = 2M_A \frac{eZ_A}{|{\bf q}_2|^2}F_A(q_{\rm 2t},q_{\rm 2l})
            \epsilon^{\mu}
            T^{(\gamma \gamma^{*} \to \pi^0 \pi^0)}_{\mu 0}\,,
\label{coulomb}
\end{equation}
where $M_A$ and $Z$ are, respectively, the mass and charge of the nucleus.
    In (\ref{coulomb}), the nuclear form factor $F_A(q_{\rm 2t},q_{\rm 2l})$,
which includes nuclear absorption, depends on transverse and
longitudinal components of the momentum transfer ${\bf q}_2$ measured
relative to the momentum ${\bf q}_1$ of the incident photon
($F_A$ is normalized to $F_A(0,0)=1$).
    $T^{(\gamma \gamma^{*} \to \pi^0 \pi^0)}_{\mu 0}$ is the tensor
component of the amplitude of the process
$\gamma (q_1) \gamma^{*}(q_2) \to \pi^0(p_1) \pi^0(p_2)$.

    From Lorentz and gauge invariances the general parameterization
for the amplitude $T^{(\gamma\gamma^{*}\to\pi^0\pi^0)}_{\mu 0}$
at $O(p^6)$ has the form
\begin{eqnarray}
\fl
  T^{(\gamma \gamma^{*} \to \pi^0 \pi^0)}_{\mu\nu} &=&
  A(s,t,u;q^2_2)\bigg( \frac{\tilde{s}}{2}g_{\mu\nu}
                -q_{2\mu}q_{1\nu} \bigg)
\nonumber \\ \fl &&
 +B(s,t,u;q^2_2)\Big[ 2\tilde{s}\Delta_{\mu}\Delta_{\nu}
       -\nu^2g_{\mu\nu}
       -2\nu\big( \Delta_{\mu}q_{1\nu}-q_{2\mu}\Delta_{\nu} \big)
                \Big]
\nonumber \\ \fl &&
 +D(s,t,u;q^2_2)\Big[
      \nu q_{2\mu}q_{2\nu}+\tilde{s}\Delta_{\mu}q_{2\nu}
     -q^2_2\big(\nu g_{\mu\nu}+2\Delta_{\mu}q_{1\nu}\big)
                \Big]\,,
\label{ggppamp}
\end{eqnarray}
where $\Delta_{\mu}=(p_1-p_2)_{\mu}$, $s=(q_1+q_2)^2=(p_1+p_2)^2$,
$\tilde{s}=s-q^2_2$, $t=(p_1-q_1)^2=(q_2-p_2)^2$,
$u=(p_2-q_1)^2=(q_2-p_1)^2$ and $\nu = t-u$.

     The differential cross section for the photoproduction of a neutral
pion pair in the Coulomb field of a nucleus is defined as
$$
    \d\sigma^{(\gamma A \to \pi\pi A)}_{\rm C} =
    \frac{\delta^{(3)}({\bf p}_1+{\bf p}_2-{\bf q}_1-{\bf q}_2)
          \delta (E_1+E_2-\varepsilon)}
         {4\varepsilon M_A (2\pi)^5 8E_1 E_2 M_A }\,\frac{1}{2}|T_{\rm C}|^2
         \d^3{\bf p}_1 \, \d^3{\bf p}_2 \, \d^3{\bf q}_2
$$
where $E_i$ and ${\bf p}_i\, (i=1,2)$ and $\varepsilon$ are the energies
and momenta of the pions and the energy of incident real photon,
respectively.
   For large energies of the incident real photon and for small momentum
transfers $|{\bf q}_2|$ to the recoil nucleus and neglecting the
offshellness of the Coulomb photon and nuclear corrections, the method of
equivalent photons \cite{Akhiezer} allows us to relate the differential
cross section for photoproduction of pion pairs on nuclei to the total
cross section for the process $\gamma \gamma \to \pi\pi$:
\begin{equation}
  {\d\sigma^{(\gamma A \to \pi\pi A)}_{\rm C} \over \d s} =
  \frac{\alpha}{\pi}Z^2 \log\bigg(\frac{\sqrt{s}}{2m_\pi}\bigg)
  \frac{1}{s} \sigma^{(\gamma \gamma \to \pi\pi)}(s)\,,
\label{eqph}
\end{equation}
where $s \equiv m^2_{\pi\pi}$.
   In this limit equation (\ref{eqph}) enables us to extract model-independent
information on the process $\gamma\gamma\to\pi\pi$ from the experimental
data on the nuclear Coulomb photoproduction of pion pairs.
   For a more general kinematics, the nuclear form factor and the
offshellness of the Coulomb photon have to be taken into account,
however.

   The nuclear form factor $F_A(q_{\rm 2t},q_{\rm 2l})$ in
(\ref{coulomb}) can be estimated in the same approximation as in
paper \cite{Faldt} since the amplitude of Coulomb photoproduction on
a single proton at small $q_{\rm 2t}$ is -- upon evaluating
(\ref{ggppamp}) -- proportional to $q_{\rm 2t}$:
$$
\epsilon^{\mu}T^{(\gamma \gamma^{*} \to \pi^0 \pi^0)}_{\mu 0} \approx
\big( \epsilon \cdot q_{\rm 2t}\big) \Big[
     A-2\Big( \tilde{s}\big(E_1-E_2\big)-\nu \varepsilon
             +2\nu \big(E_1-E_2\big)\Big) B
    -2\varepsilon D \Big]\,.
$$
   In the approach of paper \cite{Faldt} the nucleus is treated as a
completely absorbing sphere, with the form factor
\begin{equation}
\fl
F_A(q_t,q_l) = q_l R J_0(q_tR)K_1(q_lR)
              +\frac{(q_lR)^2}{q_tR} J_1(q_tR)K_0(q_lR)
              +\Delta F_A(q_t,q_l)\,,
\label{formf}
\end{equation}
where $R$ is the radius of the nucleus, and $J_n$ and $K_n$ ($n=0,1$)
are Bessel functions.
   In our estimates $R$ is chosen to be $R =1.12 A^{1/3}$ fm, where $A$
is atomic weight of a nucleus.
   The first two terms in (\ref{formf}) arise from the integration
over the three-dimensional space outside a cylinder of radius $R$ of
the nucleus.
   They reflect the assumption, that the nucleus is completely `black'
for the outgoing pions for impact parameters $b \le R$, resulting in a
cut $\theta (b-R)$, in the profile function.
   This drastic assumption is mediated by the correction term
$\Delta F_A(q_t,q_l)$, which arises from the integration over the
three-dimensional cylinder behind the nucleus and corresponds to the
interaction of photons with the nuclear Coulomb field after passing
through the nucleus.

   In general, the integrations above can be done only numerically.
   However, when both $(q_lR)$ and $(q_tR)$ are small compared to unity,
the correction $\Delta F_A(q_t,q_l)$ can conveniently be expanded
to obtain
\begin{eqnarray*}
\fl
\Delta F_A(q_t,q_l) &=& \frac{1}{4} \bigg\{
   1- q_l R J_0(q_tR)K_1(q_lR)
    -\frac{(q_lR)^2}{q_tR} J_1(q_tR)K_0(q_lR)
\\ \fl &&
  +\big[ (q_lR)^2+(q_tR)^2 \big]\bigg(
    -\frac{1}{6}+\frac{\i}{8}q_lR+\frac{1}{120}(q_tR)^2
\\ \fl &&
    +\frac{4-\i 15\pi}{480}(q_lR)^2+\i\frac{5}{144}(q_lR)^3
    -\frac{\i}{96}q_lR(q_tR)^2 \bigg)+...\bigg\} \, .
\end{eqnarray*}

\section{Chiral lagrangians}
    The calculation of the amplitude of
$\gamma\gamma^{*}\to\pi^0\pi^0$ at $O(p^6)$ of the momentum expansion
of ChPT involves tree-level, one-loop and two-loop diagrams of a chiral
effective lagrangian.
    The effective meson lagrangian at $O(p^2)$  can be written in the
nonlinear parameterization of chiral $SU(2)\times SU(2)$ symmetry as
\begin{equation}
{\cal L}^{(2)}_{\rm eff}= \frac{F^2_0}{4} \,\mbox{tr}\,
                      \big( D_{\mu}U \overline{D}^{\mu}U^\dagger \big)
                     +\frac{F_0^2}{4} \,\mbox{tr}\,
                      \big( \chi U^\dagger + U\chi^\dagger \big)\,,
\label{l2eff}
\end{equation}
   where
\begin{equation}
  U(x) = \exp \left(\frac{\i}{F_0} \varphi (x) \right)\,,\quad
  \varphi = \left ( \begin{array}{ccc}
               \pi^0        &  \sqrt{2}\pi^+ \\
            \sqrt{2}\pi^-   &      -\pi^0
                    \end{array}\right )
\nonumber
\end{equation}
represents the pseudoscalar degrees of freedom, $F_0$ is the bare
$\pi$ decay constant, and $\chi  = \mbox{diag}\,(\chi^2_u,\chi^2_d)$
is the mass matrix.
    The covariant derivatives $D_{\mu}$ and $\overline{D}_{\mu}$
contain the vector and axial-vector degrees of freedom, and are defined
as
$$
D_{\mu}U = \partial_{\mu}U + (A^L_{\mu}U - UA^R_{\mu})\,,
\quad
\overline{D}_{\mu}U^\dagger = \partial_{\mu}U^\dagger
          + (A^R_{\mu}U^\dagger - U^\dagger A^L_{\mu})\,,
$$
where $A^{R/L}_{\mu} = V_{\mu} \pm A_{\mu}$ are the right/left
combinations of vector and axial--vector fields.
    The interaction with the electromagnetic field ${\cal A}_\mu$
can be included by replacing $V_\mu \to  V_\mu + \i e {\cal A}_\mu Q$,
where Q is the matrix of quark electric charges.

    The effective meson Lagrangian of $O(p^4)$ is presented in the general
form with structure coefficients $L_i$ and $H_i$ introduced by
Gasser and Leutwyler in paper \cite{Gasser1},
\begin{eqnarray}
\fl
{\cal L}^{(4)}_{\rm eff} &=&
\bigg( L_1-\frac{1}{2}L_2 \bigg)\,
                          \big( \mbox{tr} L_{\mu} L^{\mu}\big)^2
  + L_2 \mbox{tr} \bigg(\frac{1}{2}[L_\mu,L_\nu]^2+3(L_\mu L^\mu)^2
                  \bigg)
\nonumber \\ \fl
&&+ L_3 \mbox{tr} \big( (L_{\mu} L^{\mu})^2 \big)
  - L_4 \mbox{tr} \big( L_\mu L^\mu \big)\,
        \mbox{tr} \big( \chi U^\dagger + U\chi^\dagger \big)
\nonumber \\ \fl
&&- L_5 \mbox{tr} \big[ L_\mu L^\mu \big( \chi U^\dagger
                                         + U\chi^\dagger \big)\big]
  + L_6\,\Big( \mbox{tr} \big( \chi U^\dagger + U\chi^\dagger \big)
         \Big)^2
\nonumber \\ \fl
&&+ L_7\,\Big( \mbox{tr} \big( \chi U^\dagger - U\chi^\dagger \big)
         \Big)^2
  + L_8 \mbox{tr} \big( \chi U^\dagger \chi U^\dagger
                       + U\chi^\dagger U\chi^\dagger \big)
\nonumber \\ \fl
&&- L_9 \mbox{tr} \Big( F^R_{\mu\nu}R^\mu R^\nu
                       +F^L_{\mu\nu}L^\mu L^\nu \Big)
  - L_{10} \mbox{tr} \Big( U F^R_{\mu \nu} U^\dagger F^{L\,\mu \nu}
                     \Big)
\nonumber \\ \fl
&&- H_1 \mbox{tr} \Big( \big( F^R_{\mu \nu} \big)^2
                 +\big( F^L_{\mu \nu} \big)^2 \Big)
  + H_2 \mbox{tr} \big(\chi \chi^\dagger \big)\,\,.
\label{l4eff}
\end{eqnarray}
   In equation (\ref{l4eff}), $L_{\mu}=D_{\mu}U\!\cdot \!U^\dagger$,
$R_\mu=U^\dagger\,D_\mu U$, and
$F^{R/L}_{\mu\nu} = F^V_{\mu\nu} \pm F^A_{\mu\nu}$ are the
right/left combinations of the field strength tensors
$$
F^{V}_{\mu\nu} = \partial_\mu V_\nu -\partial_\nu V_\mu
                +[V_\mu ,V_\nu ]+[A_\mu ,A_\nu ]\,,\;\;
F^{A}_{\mu\nu} = \partial_\mu A_\nu -\partial_\nu A_\mu
                +[V_\mu ,A_\nu ]+[A_\mu ,V_\nu ]\,.
$$

   It is convenient to present the part of the effective lagrangian
at $O(p^6)$ for $\gamma \gamma \to \pi^0 \pi^0$ with structure
coefficients $d_i$,
\begin{eqnarray}
\fl
{\cal L}_6&=& \frac{8}{F_0^2}\left[
 d_1 {\cal F}_{\mu\alpha} {\cal F}^{\mu\beta}
 \mbox{tr}\left(\partial^\alpha U_0
\partial_\beta U_0^\dagger Q^2\right)
+d_2 {\cal F}_{\mu\nu} {\cal F}^{\mu\nu}\mbox{tr}
 \left(\partial_\alpha U_0 \partial^\alpha U_0^\dagger Q^2\right) \right.
\nonumber\\ \fl&&
+d_3 {\cal F}_{\mu\nu} {\cal F}^{\mu\nu} \mbox{tr}
 \left(\chi (U_0+U_0^\dagger )Q^2 \right)
+d_4 {\cal F}_{\mu\nu} {\cal F}^{\mu\nu} \mbox{tr}
 (Q^2)\mbox{tr}\left(\chi (U_0+U_0^\dagger)\right)
\nonumber\\ \fl&&
+d_5 {\cal F}_{\mu\alpha} {\cal F}^{\mu\beta} \mbox{tr}\left(Q^2\right)
 \mbox{tr}\left(\partial^\alpha U_0 \partial_\beta U_0^\dagger \right)
+d_6 {\cal F}_{\mu\nu} {\cal F}^{\mu\nu} \mbox{tr}\left(Q^2\right)
 \mbox{tr}\left(\partial_\alpha U_0 \partial^\alpha U_0^\dagger \right)
\nonumber\\ \fl&&
+ d_7 {\cal F}_{\mu\alpha} {\cal F}^{\mu\beta}
\mbox{tr}\left(\partial^\alpha U_0 U_0^\dagger Q\right)
\mbox{tr}\left(\partial_\beta U_0 U_0^\dagger Q\right)
\nonumber\\ \fl&&
\left.
+d_8 {\cal F}_{\mu\nu} {\cal F}^{\mu\nu} \mbox{tr}
  \left(\partial_\alpha U_0 U_0^\dagger Q\right)
  \mbox{tr}\left(\partial^\alpha U_0 U_0^\dagger Q \right)
\right]\,,
\label{genlag}
\end{eqnarray}
   where ${\cal F}_{\mu\nu} =
        \partial_\mu {\cal A}_\nu-\partial_\nu {\cal A}_\mu$
is the standard electromagnetic field strength tensor, and
$U_0=\exp(\i\varphi_0/F_0)$, $\varphi_0=\mbox{diag}(\pi^0,-\pi^0)$.
   The lagrangian of equation (\ref{genlag}) can be obtained from the
most general representation of the full lagrangian of
paper \cite{Fearing}.

   The structure coefficients of the $O(p^4)$ and $O(p^6)$ chiral
lagrangians of equations (\ref{l4eff}) and (\ref{genlag}) can be fixed
either from low-energy meson phenomenology, as in
papers \cite{Gasser1,Bellucci1}, or from the bosonization of NJL-type
effective quark models
(see papers \cite{Bijnens2,Belkov3,Bellucci2,Belkov2} and references
therein).

\section{Amplitudes of $\gamma\gamma^{*}\to\pi^0\pi^0$}
   In general, the prediction for the Born amplitude of
$\gamma\gamma^{*}\to\pi^0\pi^0$ at $O(p^6)$ involves eight structure
coefficients $d_i$ of the lagrangian (\ref{genlag}).
   In the NJL model only the structure constants $d_1$, $d_2$, $d_3$ get
nonzero values and contribute to the amplitude:
\begin{eqnarray*}
\fl
  A^{{\rm B}(p^6)} &=& \frac{64e^2}{9F^4_0}\bigg[
       \frac{5}{16}d_1(s+q^2_2) + \frac{5}{2}d_2(s-2m^2_{\pi})
      +d_3(4\chi^2_u + \chi^2_d) \bigg]\,,
\nonumber \\ \fl
  B^{{\rm B}(p^6)} &=& -\frac{10e^2}{9F^4_0} d_1\,.
\end{eqnarray*}

     The one-loop amplitude of $\gamma\gamma^{*}\to\pi^0\pi^0$
at $O(p^4)$, which has no UV divergences, is given as
\begin{eqnarray*}
\fl
  A^{{\rm 1l}(p^4)}_{\pi} &=& -\frac{2e^2}{3F^2_0}\frac{1}{16\pi^2}
        \big( 6s-8m^2_{\pi}+\chi^2_u+\chi^2_d \big)
        \bigg[ -\frac{2q^2_2}{\tilde{s}^2}
                \Big(J^1_{\pi}(s)-J^1_{\pi}(q^2_2)\Big)
               +H_{\pi}(s,q^2_2)\bigg]\,,
\nonumber \\ \fl
  B^{{\rm 1l}(p^4)}_{\pi} &=&D^{{\rm 1l}(p^4)}_{\pi} = 0\,,
\end{eqnarray*}
    with
$$
  H_{\pi}(s,q^2_2) = \frac{1}{\tilde{s}}
             \bigg[ 1+\frac{2m^2_{\pi}}{\tilde{s}}\Big(
             J^{-1}_{\pi}(s)-J^{-1}_{\pi}(q^2_2)\Big) \bigg]\,,
$$
   and
$$
J^n_{\pi}(a) = \int^{1}_{0}d^4x\,x^n \log\bigg[
             \frac{m^2_\pi-a(1-x)x-i\epsilon }{m^2_\pi} \bigg].
$$

    In our approach UV divergences, resulting from meson loops at $O(p^6)$,
are separated using the superpropagator regularization method
\cite{Volkov2}, which is particularly well-suited for the treatment of
loops in nonlinear chiral theories.
   The result is equivalent to the dimensional regularization technique
used in paper \cite{Bellucci1}, the difference being that the scale
parameter $\mu$ is no longer arbitrary but fixed by the inherent scale
$\tilde{\mu}$ of the chiral theory, namely, $\tilde{\mu} = 4\pi \, F_0$.
   For comparison of these two methods, the constants from the UV
divergences in the dimensional regularization have to be replaced
by a finite term using the substitution
$$
(C - 1/\varepsilon) \quad \longrightarrow \quad C_{\rm SP}
=2C+1+\frac12\,\left[{\d \over \d z}
                     \left(\log \Gamma^{-2}(2z+2)
                     \right)
               \right]_{z=0} + \beta \pi
=-1+4C + \beta \pi\,,
$$
   where $C=0.577$ is Euler's constant, $\varepsilon=(4-D)/2$, and $\beta$
is an arbitrary constant resulting from the representation of the
superpropagator as an integral of the Sommerfeld-Watson type.
   The splitting of the decay constants $F_\pi$ and $F_K$ is used
at $O(p^4)$ to fix $C_{\rm SP} \approx 3.0$ for $F_0 = 90$~MeV.

    The one-loop diagrams at $O(p^6)$ involve the structure coefficients
$L_i$ of the lagrangian (\ref{l4eff}) and give the following
contributions
\begin{eqnarray*}
\fl
A^{{\rm 1l}(p^6)}_{\pi} &=&-\frac{e^2}{3F^4_0}\frac{1}{16\pi^2}\bigg\{\bigg[
      2L_9q^2_2\big( 6s-8m^2_{\pi}+\chi^2_u+\chi^2_d \big)
\nonumber \\ \fl &&
     +12L_2\Big(3s^2-8m^2_{\pi}\big(s-m^2_{\pi}\big)\Big)
     +24\big(2L_1-L_2+L_3\big)\big(s-2m^2_{\pi}\big)^2
\nonumber \\ \fl &&
     +16L_4\Big( 3\big(s-2m^2_{\pi}\big)\big(\chi^2_u+\chi^2_d \big)
                -\big(3s-4m^2_{\pi}\big)\mbox{tr}\chi \Big)
\nonumber \\ \fl &&
     -16L_5m^2_{\pi}(\chi^2_u+\chi^2_d )\bigg]  
        2\bigg[-\frac{2q^2_2}{\tilde{s}^2}
                \Big(J^1_{\pi}(s)-J^1_{\pi}(q^2_2)\Big)
               +H_{\pi}(s,q^2_2)\bigg]
\nonumber \\ \fl &&
    +4\big(L_9+L_{10}\big)\big(6s-8m^2_{\pi}+\chi^2_u+\chi^2_d\big)
     G_{\pi}(s)
\nonumber \\ \fl &&
    +12L_2\bigg[ \frac{2}{3}J^1_{\pi}(s)\bigg( s-2m^2_{\pi}
    -\frac{1}{4\tilde{s}}\big(3q^4_2-48q^2_2m^2_{\pi}+32m^4_{\pi}\big)
\nonumber \\ \fl &&~~~~~~~~~~~~~~~~~~
    -\frac{1}{2\tilde{s}^2}\Big(80q^2_2m^4_{\pi}-21q^4_2m^2_{\pi}
                       +2q^6_2+6\nu^2\big(q^2_2+12m^2_{\pi}\big) \Big)
\nonumber \\ \fl &&~~~~~~~~~~~~~~~~~~
    -\frac{6}{\tilde{s}^3}\nu^2q^2_2\big(q^2_2+19m^2_{\pi}\big)
    -\frac{3}{\tilde{s}^4}\nu^2q^4_2\big(q^2_2+26m^2_{\pi}\big) \bigg)
\nonumber \\ \fl &&~~~~~~~~
    +\frac{2}{3}J^1_{\pi}(q^2_2)\bigg(4m^4_{\pi}-q^4_2
    +\frac{q^2_2}{\tilde{s}}\big(q^2_2-10m^2_{\pi}\big)
\nonumber \\ \fl &&~~~~~~~~~~~~~~~~~~
    +\frac{1}{\tilde{s}^2}\Big(40q^2_2m^4_{\pi}-14q^4_2m^2_{\pi}
                       +q^6_2+3\nu^2\big(q^2_2+4m^2_{\pi}\big) \Big)
\nonumber \\ \fl &&~~~~~~~~~~~~~~~~~~
    +\frac{6}{\tilde{s}^3}\nu^2q^2_2\big(q^2_2+14m^2_{\pi}\big)
    +\frac{3}{\tilde{s}^4}\nu^2q^4_2\big(q^2_2+26m^2_{\pi}\big) \bigg)
\nonumber \\ \fl &&~~~~~~~~
    -2H_{\pi}(s,q^2_2)\bigg( \tilde{s}m^2_{\pi}-4m^4_{\pi}
    +q^2_2m^2_{\pi}-\nu^2
    -\frac{2}{\tilde{s}}\nu^2\big(m^2_{\pi}+2q^2_2\big)
\nonumber \\ \fl &&~~~~~~~~~~~~~~~~~~
    -\frac{3}{\tilde{s}^2}\nu^2q^2_2\big(m^2_{\pi}+q^2_2\big)\bigg)
\nonumber \\ \fl &&~~~~~~~~
    -\frac{1}{18}\big(s-4m^2_{\pi}\big)
    +\frac{1}{3\tilde{s}}\big(q^4_2-14q^2_2m^2_{\pi}-5\nu^2\big)
    -\frac{13}{2\tilde{s}^2}\nu^2q^2_2
\nonumber \\ \fl &&~~~~~~~~
    -\frac{5}{\tilde{s}^3}\nu^2q^4_2
    +\frac{1}{3}\widetilde{C}_{\pi}\big(\tilde{s}-4m^2_{\pi}\big)
     \bigg] \bigg\}\,,
\nonumber \\ \fl
  B^{{\rm 1l}(p^6)}_{\pi} &=& -\frac{4e^2}{F^4_0}\frac{1}{16\pi^2}L_2\bigg\{
        \frac{1}{3}J^1_{\pi}(s)\bigg[1
       -\frac{2}{\tilde{s}}\big(5m^2_{\pi}-q^2_2\big)
       -\frac{q^2_2}{\tilde{s}^2}\big(10m^2_{\pi}-q^2_2\big)\bigg]
\nonumber \\ \fl &&
       +\frac{1}{3\tilde{s}}J^1_{\pi}(q^2_2)\bigg[
        2\big(4m^2_{\pi}-q^2_2\big)
       +\frac{q^2_2}{\tilde{s}}\big(10m^2_{\pi}-q^2_2\big)\bigg]
\nonumber \\ \fl &&
       +m^2_{\pi}H_{\pi}(s,q^2_2)-\frac{q^2_2}{6\tilde{s}}
       -\frac{7}{36}+\frac{1}{6}\widetilde{C}_{\pi} \bigg\}\,,
\nonumber \\ \fl
  D^{{\rm 1l}(p^6)}_{\pi} &=& \frac{8e^2}{3F^4_0}\frac{1}{16\pi^2}L_2
                        \frac{\nu}{\tilde{s}^3}\bigg\{
        q^2_2 J^1_{\pi}(s)\Big[\tilde{s}^2
       +2\tilde{s}\big(13m^2_{\pi}+q^2_2\big)
       +\big(26m^2_{\pi}+q^2_2\big)\Big]
\nonumber \\ \fl &&
       +J^1_{\pi}(q^2_2)\Big[
        \tilde{s}^2\big(4m^2_{\pi}-q^2_2\big)
       -2\tilde{s}q^2_2\big(8m^2_{\pi}+q^2_2\big)
       -q^4_2\big(26m^2_{\pi}+q^2_2\big)\Big]
\nonumber \\ \fl &&
       -3q^2_2H_{\pi}(s,q^2_2)\tilde{s}^2\big(s+m^2_{\pi}\big)
       +\frac{1}{4}q^2_2\tilde{s}\big(9s+q^2_2\big) \bigg\}\,,
\end{eqnarray*}
where $G_{\pi}(s) = \widetilde{C}_{\pi}+2J^1_{\pi}(s)$ and
$\widetilde{C}_{\pi} = C_{\rm SP}+\log\frac{m^2_\pi}{16\pi^2F^2_0}$.

    Only two-loop diagrams which are factorizable can be calculated
analytically.
    Their contribution can be presented in the form
\begin{eqnarray*}
\fl
A^{{\rm 2l}(p^6)}_{\pi} &=&-\frac{2e^2}{3F^4_0}\frac{1}{(16\pi^2)^2}\bigg\{
     J^1_{\pi}(q^2_2)\bigg( \frac{q^2_2}{\tilde{s}^2}J^1_{\pi}(q^2_2)
                           +\frac{1}{2}H_{\pi}(s,q^2_2)\bigg)
     \frac{1}{\tilde{s}^3}\bigg[ -\frac{32}{3}m^4_{\pi}\tilde{s}^3
\nonumber \\ \fl&&~~~~~~~
    +2s^4\big(4m^2_{\pi}-q^2_2\big)
    -\frac{2}{3}s^3q^2_2\big(32m^2_{\pi}-9q^2_2\big)
    +2s^2q^4_2\big(8m^2_{\pi}-3q^2_2\big)
\nonumber \\ \fl&&~~~~~~~
    -2sq^4_2\big(32m^4_{\pi}-q^4_2\big)
    -\frac{8}{3}m^2_{\pi}q^8_2
    +\frac{1}{3}\tilde{s}^3\big(4m^2_{\pi}-q^2_2\big)
                   \big(\chi^2_u+\chi^2_d\big)\bigg]
\nonumber \\ \fl&&
    -\frac{1}{3\tilde{s}^5}J^1_{\pi}(q^2_2)\bigg[
     6s^5\big(4m^2_{\pi}-q^2_2\big)
    -s^4\big(20m^4_{\pi}+15m^2_{\pi}q^2_2-29q^4_4\big)
\nonumber \\ \fl&&~~~~~~~
    -16m^2_{\pi}\tilde{s}^4
    -\frac{2}{3}s^3q^2_2\big(16m^4_{\pi}+166m^2_{\pi}q^2_2+9q^4_2\big)
\nonumber \\ \fl&&~~~~~~~
    +2s^2q^4_2\big(60m^4_{\pi}+91m^2_{\pi}q^2_2+15q^4_2\big)
\nonumber \\ \fl&&~~~~~~~
    -2sq^6_2\big(176m^4_{\pi}+46m^2_{\pi}q^2_2+q^4_2\big)
    +\frac{1}{3}m^2_{\pi}q^8_2\big(276m^2_{\pi}+35q^2_2\big)
\nonumber \\ \fl&&~~~~~~~
    +\frac{1}{6}\bigg(~
     3\big(\tilde{s}^4+s^4\big)\big(4m^2_{\pi}-q^2_2\big)
    -3s^3q^2_2\big(36m^2_{\pi}-7q^2_2\big)
\nonumber \\ \fl&&~~~~~~~~~~~~~
    -24s^2q^4_2\big(6m^2_{\pi}-q^2_2\big)
    +6sq^6_2\big(20m^2_{\pi}-3q^2_2\big)
\nonumber \\ \fl&&~~~~~~~~~~~~~
    +q^8_2\big(36m^2_{\pi}-5q^2_2\big)\bigg)
          \big(\chi^2_u+\chi^2_d\big) \bigg]
\nonumber \\ \fl&&
    +J^1_{\pi}(s)\bigg( \frac{q^2_2}{\tilde{s}^2}J^1_{\pi}(s)
                           -\frac{1}{2}H_{\pi}(s,q^2_2)\bigg)
     \frac{1}{\tilde{s}^3}\bigg[ \frac{64}{3}m^4_{\pi}\tilde{s}^3
    +12s^5
\nonumber \\ \fl&&~~~~~~~
    -4s^4\big(8m^2_{\pi}+9q^2_2\big)
    +\frac{4}{3}s^3q^2_2\big(32m^2_{\pi}+9q^2_2\big)
    -12s^2q^4_2\big(8m^2_{\pi}+q^2_2\big)
\nonumber \\ \fl&&~~~~~~~
    +32sm^4_{\pi}q^6_2
    +\frac{16}{3}\tilde{s}^3\big(3s-4m^2_{\pi}\big)
                 \big(\chi^2_u+\chi^2_d\big)
    +\frac{7}{3}\tilde{s}^3\big(\chi^2_u+\chi^2_d\big)^2\bigg]
\nonumber \\ \fl&&
    +\frac{1}{3\tilde{s}^5}J^1_{\pi}(s)\bigg[
     6s^5m^2_{\pi}
    -2s^4\big(20m^4_{\pi}+88m^2_{\pi}q^2_2+19q^4_2\big)
    +32m^4_{\pi}\tilde{s}^4
\nonumber \\ \fl&&~~~~~~~
    +2s^3q^2_2\bigg(136m^4_{\pi}-\frac{502}{3}m^2_{\pi}q^2_2-5q^4_2
              \bigg)
\nonumber \\ \fl&&~~~~~~~
    -6s^2q^4_2\bigg(96m^4_{\pi}+\frac{124}{3}m^2_{\pi}q^2_2+q^4_2
              \bigg)
\nonumber \\ \fl&&~~~~~~~
    +2sq^6_2\big(248m^4_{\pi}+43m^2_{\pi}q^2_2+q^4_2\big)
    -\frac{8}{3}m^2_{\pi}q^8_2\big(57m^2_{\pi}+q^2_2\big)
\nonumber \\ \fl&&~~~~~~~
    +\frac{1}{3}\bigg(-72s^5+36s^4\big(3m^2_{\pi}+8q^2_2\big)
    +24\tilde{s}^4\big(3s-4m^2_{\pi}\big)
\nonumber \\ \fl&&~~~~~~~~~~~~~
    -s^3q^2_2\big(420m^2_{\pi}+433q^2_2\big)
    +3s^2q^4_2\big(204m^2_{\pi}+97q^2_2\big)
\nonumber \\ \fl&&~~~~~~~~~~~~~
    -3sq^6_2\big(132m^2_{\pi}+25q^2_2\big)
    +q^8_2\big(96m^2_{\pi}+q^2_2\big)\bigg)
          \big(\chi^2_u+\chi^2_d\big) \bigg]
\nonumber \\ \fl&&
    +\frac{q^2_2}{\tilde{s}^5}J^1_{\pi}(q^2_2)J^1_{\pi}(s)\bigg[
    -12s^5+2s^4\big(12m^2_{\pi}+19q^2_2\big)
    -\frac{32}{3}m^4_{\pi}\tilde{s}^3
\nonumber \\ \fl&&~~~~~~~
    -\frac{14}{3}s^3q^2_2\big(16m^2_{\pi}+3q^2_2\big)
    +2s^2q^4_2\big(40m^2_{\pi}+9q^2_2\big)
\nonumber \\ \fl&&~~~~~~~
    -2sq^6_2\big(16m^2_{\pi}+q^2_2\big)+\frac{8}{3}m^2_{\pi}q^8_2
    +\frac{1}{3}\bigg( 20m^2_{\pi}\tilde{s}^3-48s^4+145s^3q^2_2
\nonumber \\ \fl&&~~~~~~~
    -147s^2q^4_2+51sq^6_2-q^8_2\bigg)\big(\chi^2_u+\chi^2_d\big)
    -\frac{7}{3}\tilde{s}^3\big(\chi^2_u+\chi^2_d\big)^2\bigg]
\nonumber \\ \fl&&
    -\frac{1}{2\tilde{s}^3}H_{\pi}(s,q^2_2)\bigg[
    -\frac{2}{3}s^4\big(36m^2_{\pi}+q^2_2\big)
    +\frac{2}{9}s^3q^2_2\big(328m^2_{\pi}+9q^2_2\big)
\nonumber \\ \fl&&~~~~~~~
    +\frac{112}{3}m^4_{\pi}\tilde{s}^3
    -2s^2q^4_2\big(38m^2_{\pi}+q^2_2\big)
    +\frac{2}{3}sq^6_2\big(40m^2_{\pi}+q^2_2\big)
\nonumber \\ \fl&&~~~~~~~
    -\frac{8}{9}m^2_{\pi}q^8_2
    +\frac{1}{9}\tilde{s}^3\big(12m^2_{\pi}-q^2_2\big)
                \big(\chi^2_u+\chi^2_d\big) \bigg]
\nonumber \\ \fl&&
    -\frac{1}{9\tilde{s}^5}\bigg(s^4q^2_2\big(109m^2_{\pi}+3q^2_2\big)
    -48m^4_{\pi}\tilde{s}^4+2m^2_{\pi}s^3q^4_2-4m^2_{\pi}sq^8_2
    -3m^2_{\pi}q^{10}_2\bigg)
\nonumber \\ \fl&&
    +\widetilde{C}_{\pi}\bigg(
     \frac{q^2_2}{\tilde{s}^2}\big(J^1_{\pi}(s)-J^1_{\pi}(q^2_2)\big)
    -\frac{1}{2}H_{\pi}(s,q^2_2)\bigg)\frac{1}{\tilde{s}^3}\bigg[
     6s^5+s^4\big(8m^2_{\pi}-17q^2_2\big)
\nonumber \\ \fl&&~~~~~~~
    -\frac{1}{3}s^3q^2_2\big(76m^2_{\pi}-45q^2_2\big)
    -\frac{80}{3}m^4_{\pi}\tilde{s}^3
    +s^2q^4_2\big(28m^2_{\pi}-q^2_2\big)
\nonumber \\ \fl&&~~~~~~~
    -sq^6_2\big(12m^2_{\pi}+q^2_2\big)+\frac{4}{3}m^2_{\pi}q^8_2
    -\frac{1}{6}\big(72m^2_{\pi}\tilde{s}^3-48s^4+143s^3q^2_2
\nonumber \\ \fl&&~~~~~~~
                     -141s^2q^4_2+45sq^6_2+q^8_2\big)
                \big(\chi^2_u+\chi^2_d\big)
    +\frac{7}{6}\tilde{s}^3\big(\chi^2_u+\chi^2_d\big)^2\bigg]
\nonumber \\ \fl&&
    +\frac{1}{\tilde{s}^5}\widetilde{C}_{\pi}\bigg[
     s^5m^4_{\pi}-\frac{23}{6}s^4m^2_{\pi}q^2_2+4m^4_{\pi}\tilde{s}^4
    +\frac{1}{3}s^3q^4_2\big(3m^2_{\pi}+16q^2_2\big)
\nonumber \\ \fl&&~~~~~~~
    -s^2q^6_2\big(20m^2_{\pi}+q^2_2\big)
    +\frac{1}{3}sq^8_2m^2_{\pi}+\frac{1}{6}m^2_{\pi}q^{10}_2
\nonumber \\ \fl&&~~~~~~~
    +\frac{2}{3}m^2_{\pi}\tilde{s}^4\big(\chi^2_u+\chi^2_d\big)
     \bigg] \bigg\}\,,
\nonumber \\ \fl
  B^{{\rm 2l}(p^6)}_{\pi} &=&D^{{\rm 2l}(p^6)}_{\pi} = 0\,.
\end{eqnarray*}
    It should be stressed that additional two-loop diagrams, such as
box diagrams and acnode graphs, which cannot be evaluated analytically,
can be neglected: the numerical estimates in
paper \cite{Bellucci1} indicate the smallness of their contributions
in the photoproduction process under consideration.

\section{Numerical estimates and conclusion}
    Monte Carlo techniques were used to compute the total cross
sections for the photoproduction of $\pi^0 \pi^0$-pairs in the Coulomb
field of the carbon ($Z=6$) and silicon nuclei ($Z=14$).
    It is important to note that for a momentum transfer
cutoff $q_{\rm max} \equiv  |{\bf q}_2|_{\rm max} \ll \varepsilon$, the
effective mass of $\pi\pi$ system varies in the range
$4 m^2_\pi \leq m^2_{\pi\pi} \leq 2 \varepsilon q_{\rm max}$.

    The dependence of the total cross section of the reaction
$\gamma A \to \pi^0 \pi^0 A$ on the momentum transfer cutoff
$q_{\rm max}$ for different energies $\varepsilon$ of the incident real
photon is shown in figures 2 and 3.
   (In the calculation, the additional cutoff $m_{\pi\pi} \le 700$~MeV
was used, corresponding to the range of validity of the chiral theory.)
    In figure 2 we demonstrate explicitly the influence of the nuclear
form factor: our numerical estimates indicate that the nuclear effects
are mainly saturated by the contribution of the two first terms of
equation (\ref{formf}), while the correction $\Delta F_A$ does not exceed
10\%.
    For our numerical estimates we used for the parameters
$L_i$ and $d_i$ the values from (14) and Table~1 of
paper \cite{Belkov2} obtained from the NJL model.
   These values include the resonance-exchange contributions effectively
taken into account by integrating out the vector, axial-vector and scalar
degrees of freedom.
   Moreover, for numerical comparison with the phenomenological
approach to $\gamma\gamma\to\pi^0\pi^0$, we used the structure constants
$L_i$ and $d_i$ corresponding to Tables~1 and 2 of paper \cite{Bellucci1}.
   The results of our calculations with the parameters of
paper \cite{Bellucci1} are also shown in figure 2.

   The main background for the Coulomb photoproduction of $\pi^0 \pi^0$
pairs is the double pion photoproduction reaction on the nucleon
$\gamma N \to \pi^0 \pi^0 N$ via the baryon resonance exchange mechanism 
\cite{Luke,Gomez1}.
   In the photon energy region near $\varepsilon = 1$ GeV the total cross 
sections of the reactions $\gamma p \to \pi^0 \pi^0 p$ and 
$\gamma n \to \pi^0 \pi^0 n$ are dominated by diagrams in figure 4 with 
$\Delta(1232)$-isobar and resonance $N^{*}(1520)$ \cite{Gomez2}.
   Kroll-Ruderman diagram dominating in other isospin channels and meson
exchange diagrams do not contribute to the photoproduction of $\pi^0 \pi^0$
pairs.
   The contributions of diagrams j, l, and n of figure 4 to the reaction
$\gamma A \to \pi^0 \pi^0 A$ can be canceled by choosing an isoscalar target
because of opposite signs of the corresponding amplitudes on the proton and 
neutron. 
   The amplitude of diagram p in figure 4 is proportional to the difference
of the coupling constants $(\tilde{g}^N_\gamma - \tilde{g}^N_\sigma)$ 
(see paper \cite{Gomez2} for more details).
   The corresponding contribution to the process $\gamma A \to \pi^0 \pi^0 A$
on the isoscalar target proves to be also suppressed due to the fact that
\begin{eqnarray*}
(\tilde{g}^N_\gamma - \tilde{g}^N_\sigma) = \left\{
\begin{array}{l@{,}l}
~~0.157\;\;& \mbox{~for~proton} \,,\\
 -0.136\;\;& \mbox{~for~neutron} \,.
\end{array} \right.
\end{eqnarray*}
   The effect of such self-cancellation of diagrams j, l, n and p from figure 
4 is demonstrated in figure 5 where we present the results of our calculation
of the total cross section of the double pion photoproduction reaction on the
proton, neutron and on the system (proton + neutron).
   Numerically our calculations are in a good agreement with the results of
paper \cite{Gomez2} and in a reasonable agreement with the recent experimental 
data \cite{DAFNE2}. 

   In figure 6 we also show the photon energy dependence of the total cross 
sections for the coherent double pion photoproduction on the carbon and silicon
nuclei (isoscalar targets) calculated with the Born nuclear electromagnetic 
form factor with symmetrized Fermi-density \cite{Burov}:
$$
  F^{SF}_B(q) = 
 -\,\frac{4\pi^2 b R \rho^{(0)}}{q\,\mbox{sinh}\pi bq}
  \bigg( \mbox{cos}\,qR 
        -\frac{\pi b}{R}\mbox{sin}\,qR\,\,\mbox{cot}\,\pi bq
  \bigg)\,,
$$
where $q=|{\bf q}_1-{\bf p}_1-{\bf p}_2|$.
   The normalization constant is defined as 
$$
\rho^{(0)} = \frac{3}{4\pi}\,\frac{1}{R(R^2+\pi^2b^2)}\,,
$$
and the following values of radius of the nucleus $R$ and impact parameter
$b$ were used: $R = 2.214$ fm, $b=0.488$ fm --- for carbon, and $R = 3.085$ fm,
$b=0.563$ fm --- for silicon.
   Our results for $\sigma_{tot}(C^{12})$ agree with recent estimates 
\cite{Oset}.

   To estimate the contribution of the reaction $\gamma N \to \pi^0 \pi^0 N$
to this part of the phase space which corresponds to the cutoffs introduced 
above for the Coulomb photoproduction of $\pi^0 \pi^0$ pairs, the additional 
cutoffs must be also taken into account when Monte Carlo simulating the 
background: 
$m_{\pi \pi} \le 700$ MeV and $\sqrt{-t_N} \le q_{max}$, where 
$t_N=(p_1-p_2)^2$, $p_1$ and $p_2$ are the momenta of nucleon in the initial 
and final states, respectively.
   The photon energy dependence of the cross sections for the carbon and 
silicon targets calculated with the cutoffs are shown in figure 6.
   Our estimates of the $\gamma N \to \pi^0 \pi^0 N$ background have shown that
in the case of an isoscalar target it is strongly suppressed in the energy and 
momentum transfer regions under discussion.
   According to the symmetry properties of the background coherent reaction
$\gamma A \to \pi^0 \pi^0 A$ on isoscalar nuclei \cite{Oset}, the two $\pi^0$ 
mesons prefer to propagate together in the CMS.
   The Monte-Carlo simulation shows that the additional suppression of the
backgraund to the level lower than the Coulomb photoproduction can be easily 
achieved using some angle cutoffs. 
   To extract the signal of Coulomb photoproduction is possible in the 
experiment where a silicon detector is used as a sensitive target to develop 
a trigger on the nucleus recoil momentum.

   Summarizing, we have found that for energies of the incident photon
$\varepsilon \le 4$ GeV and for momentum transfers
$|{\bf q}_2| \le 200$ MeV, $\pi^0 \pi^0$ pairs with
$2m_\pi \le m_{\pi\pi} \le 700$MeV in the reaction
$\gamma A \to \pi^0 \pi^0 A$ are produced with a cross section of
typically $\sigma \approx 40$ pb and $\sigma \approx 120$ pb for carbon
and silicon nuclei, respectively.
   Our results demonstrate that this reaction can be experimentally
investigated with presently available photon beams as a new source
of the low-energy data on the process $\gamma \gamma \to \pi^0 \pi^0$.

\ack

   The authors gratefully acknowledge fruitful discussions with
G~Anton, V~V~Burov, S~S~Kamalov, V~K~Lukyanov, G~I~Lykasov, B~Mayer, E~Oset 
and A~V~Tarasov.

\newpage
\section*{References}

\newpage
\section*{Figures}

\begin{figure}
\hspace{23mm}\epsfbox{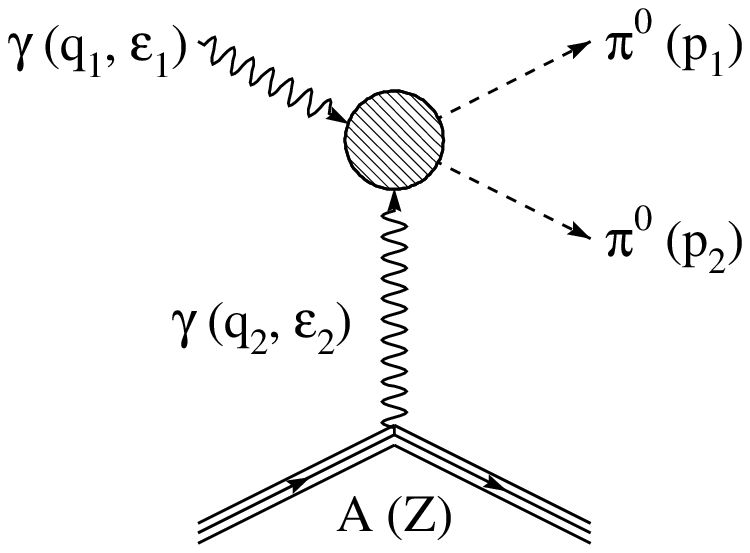}
\begin{center}
\small
{\bf Figure 1.}
   Photoproduction of pion pairs in the Coulomb field of a nucleus.
\end{center}
\end{figure}

\begin{figure}
\epsfxsize=125mm
\hspace{10mm}\epsfbox{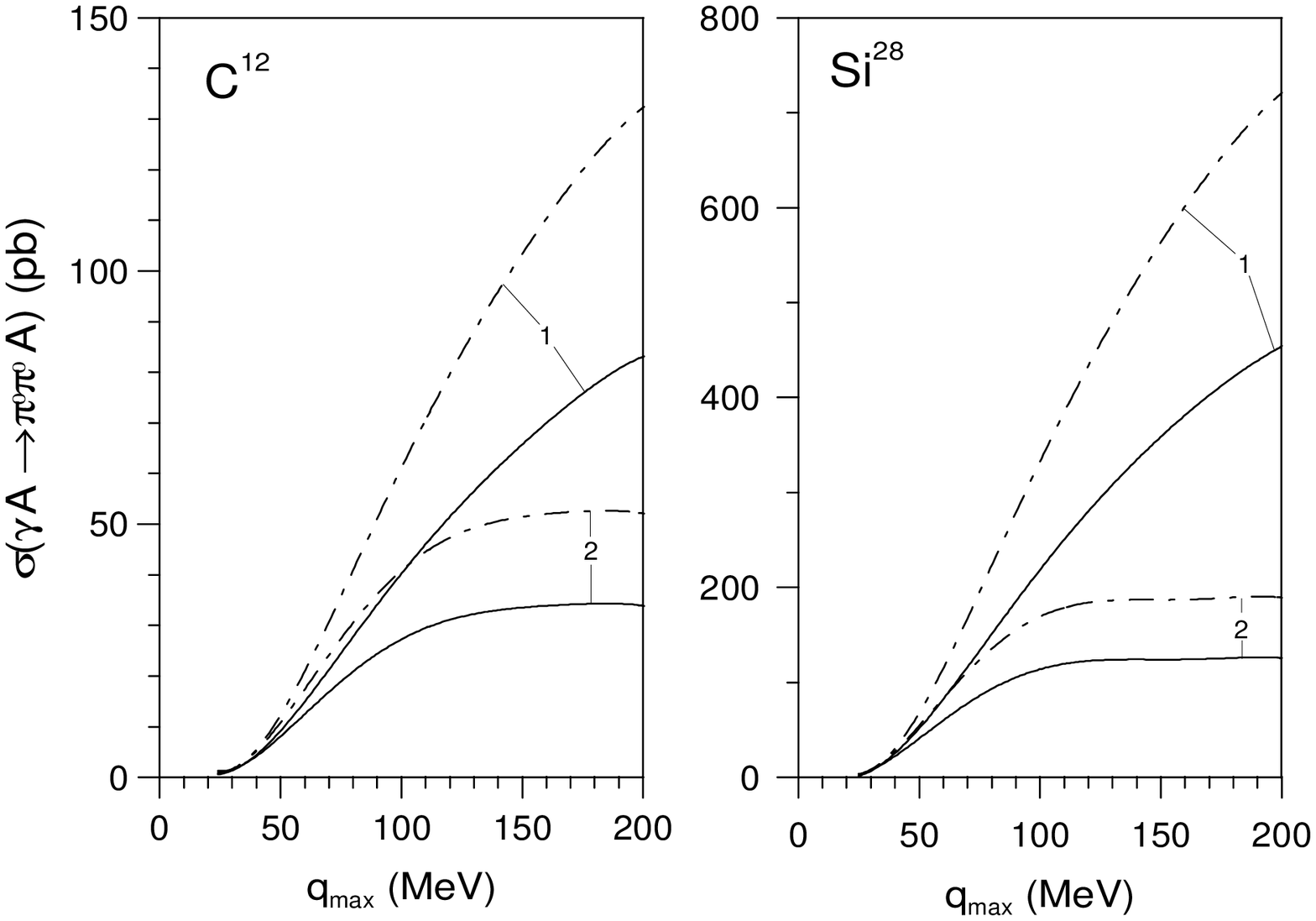}\\
\small
\noindent {\bf Figure 2.}
   Dependence of the $\gamma A \to \pi^0 \pi^0 A$ cross section on the
momentum transfer cutoff $q_{\rm max}$ for the energy $\varepsilon =4$~GeV
of the incident photon.
   Compared are: $L_i$, $d_i$ from the bosonization of the NJL model
(\cite{Belkov2}, full curves) and from phenomenology
(\cite{Bellucci1}, dash-dotted curves), without and with a nuclear form
factor (indicated by (1) and (2)).
\end{figure}

\begin{figure}
\epsfxsize=125mm
\hspace{10mm}\epsfbox{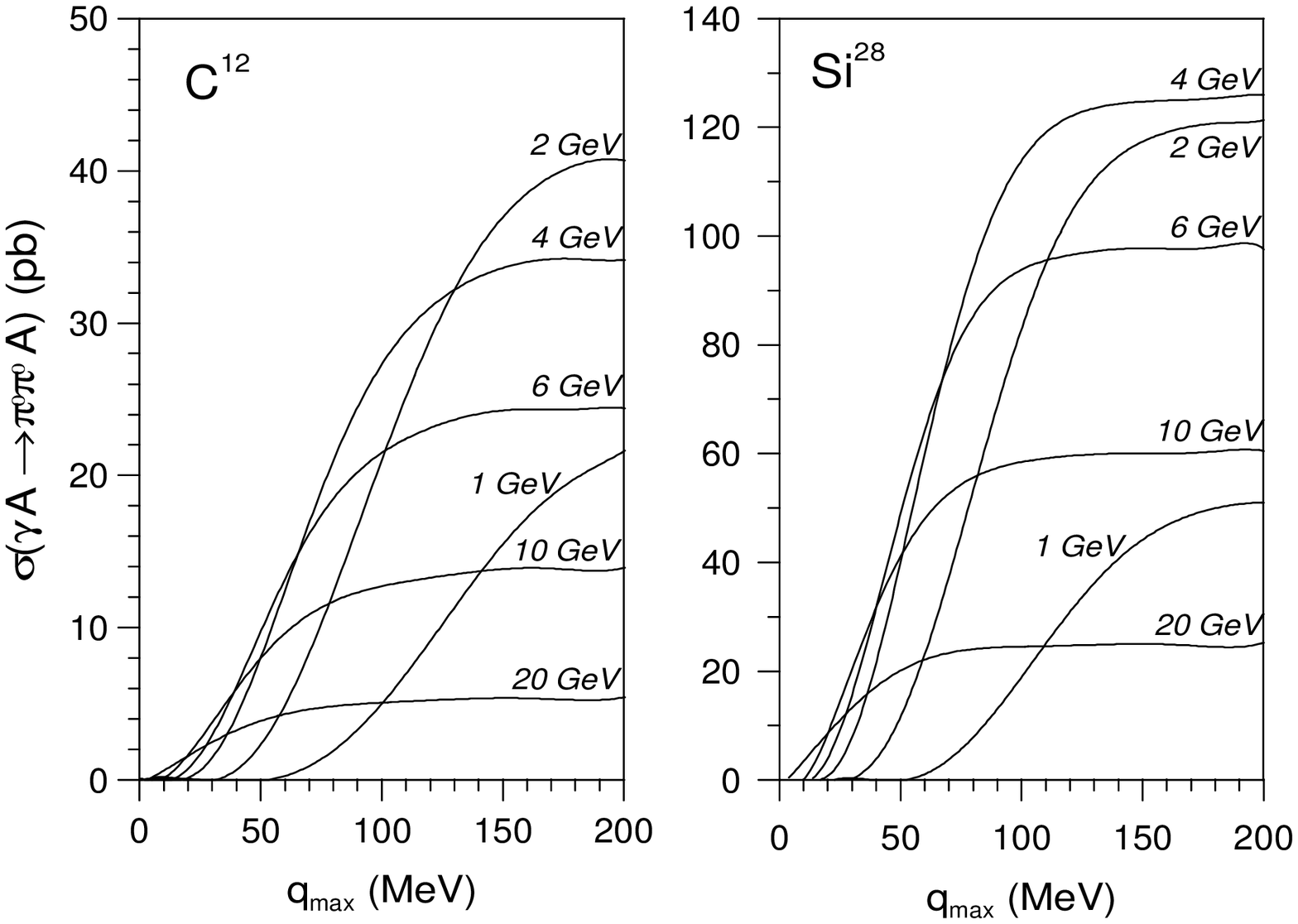}\\
\small
\noindent {\bf Figure 3.}
   Dependence of the $\gamma A \to \pi^0 \pi^0 A$ cross section on the
momentum transfer cutoff $q_{\rm max}$ for different photon
energies, with parameters $L_i$ and $d_i$ from the NJL model
\cite{Belkov2}.
\end{figure}

\begin{figure}
\epsfxsize=125mm
\hspace{10mm}\epsfbox{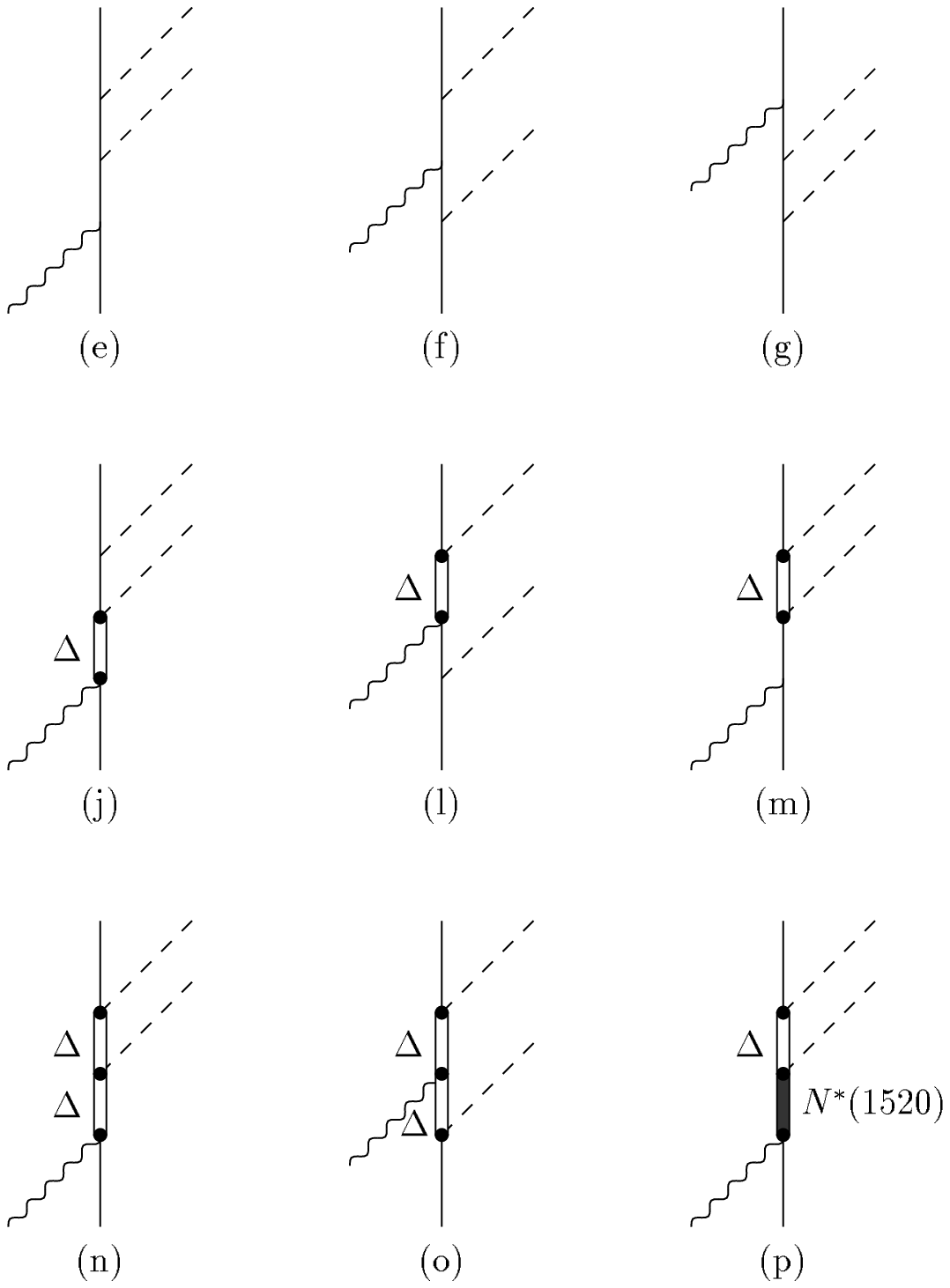}\\
\small
\noindent {\bf Figure 4.}
   Dominating Feynman diagrams for the $\gamma p \to \pi^0 \pi^0 p$ and 
$\gamma n \to \pi^0 \pi^0 n$. 
\end{figure}

\begin{figure}
\epsfxsize=125mm
\hspace{10mm}\epsfbox{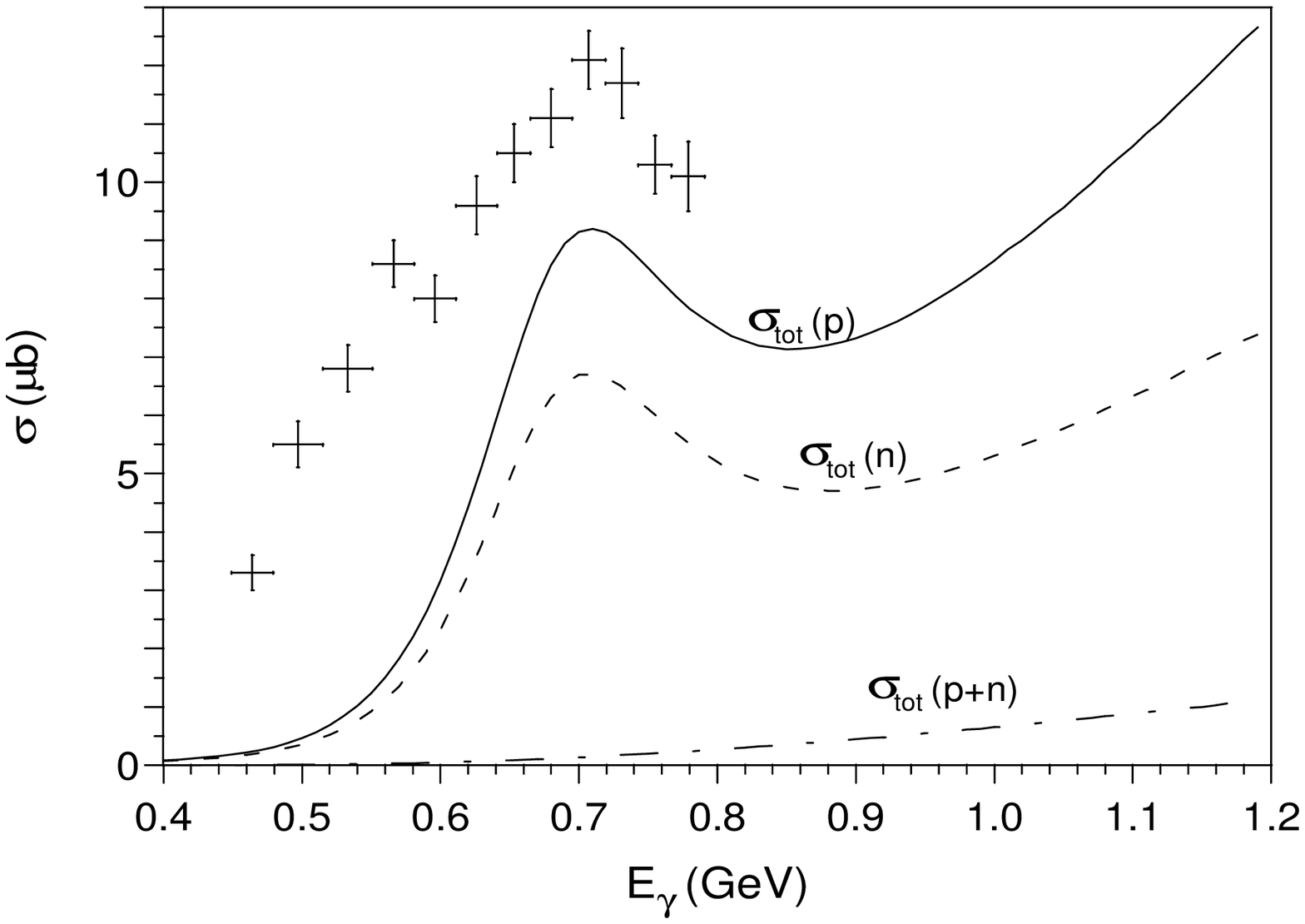}\\
\small
\noindent {\bf Figure 5.}
   Total cross sections of the reactions $\gamma p \to \pi^0 \pi^0 p$ 
(full curve) and $\gamma n \to \pi^0 \pi^0 n$ (dashed curve) calculated 
in the framework of baryon resonance model \cite{Gomez2} with diagrams 
of figure 4. Dash-dotted curve corresponds to the total cross section for the 
sum of the amplitudes on the proton and neutron. The experimental 
points for the reaction $\gamma p \to \pi^0 \pi^0 p$ were measured at 
DA$\Phi$NE \cite{DAFNE2}. 
\end{figure}

\begin{figure}
\epsfxsize=125mm
\hspace{10mm}\epsfbox{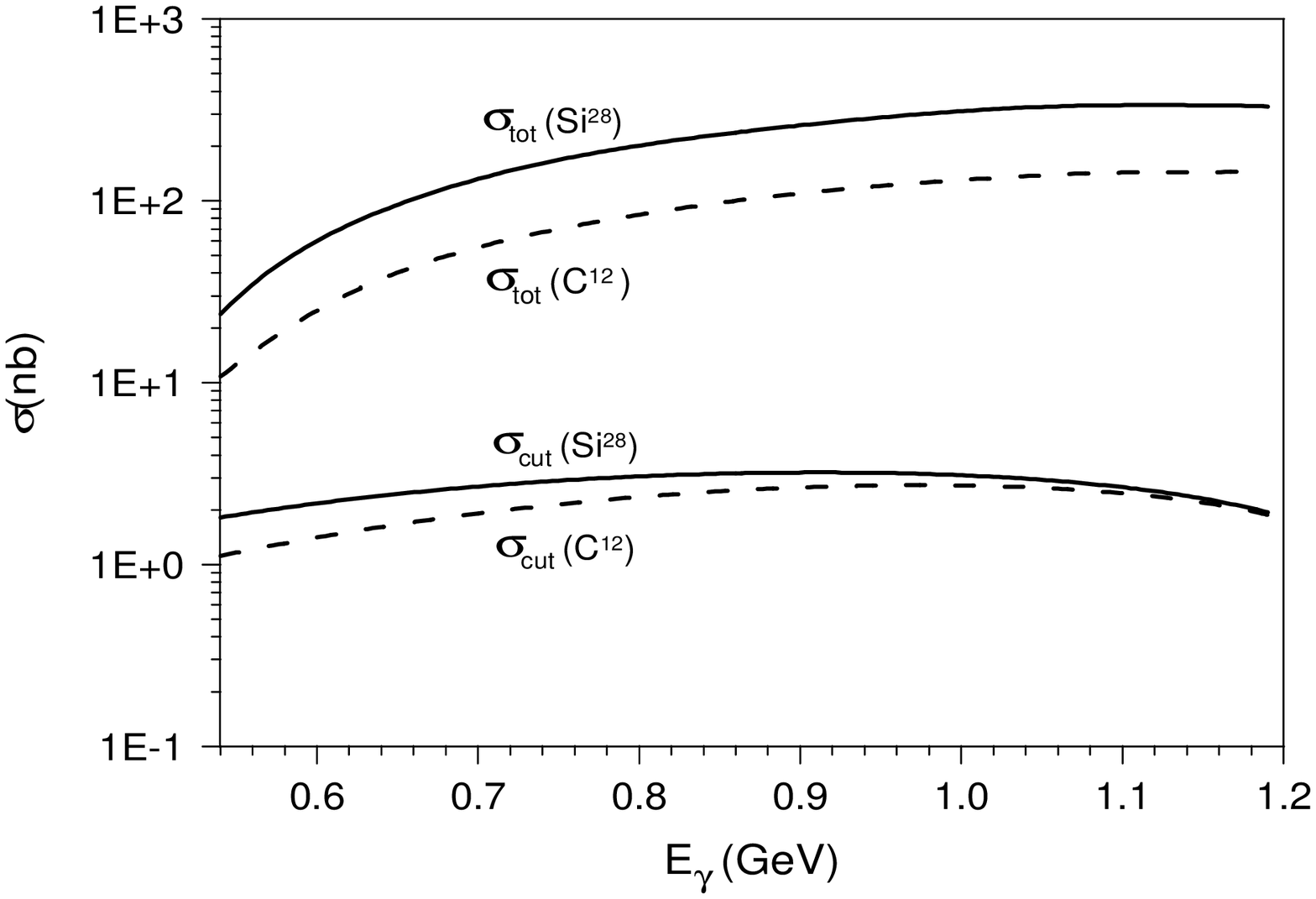}\\
\small
\noindent {\bf Figure 6.}
   Total cross sections of the double pion photoproduction reaction 
($\sigma_{tot}$) and cross sections calculated with the cutoffs 
$\sqrt{-t_N} \le 150$ MeV and $m_{\pi \pi} \le 700$ MeV ($\sigma_{cut})$ 
for coherent processes on the carbon (dashed curve) and silicon nuclei 
(full curve).
\end{figure}
\end{document}